\documentclass[preprint,aps,superscriptaddress]{revtex4}
\usepackage{graphicx}
\usepackage{amsmath}
\usepackage[usenames]{color}
\usepackage{amssymb}
\newcommand{\be}{\begin{equation}}
\newcommand{\ee}{\end{equation}}
\newcommand{\bea}{\begin{eqnarray}}
\newcommand{\eea}{\end{eqnarray}}
\newcommand{\pr}{\partial}
\newcommand{\nno}{\nonumber}
\newcommand{\bse}{\begin{subequations}}
\newcommand{\ese}{\end{subequations}}

\begin{document}
\title{Analysing the Effect on CMB in a Parity and Charge Parity 
Violating Varying Alpha Theory  }
\author{Debaprasad Maity
\footnote{debu.imsc@gmail.com}}
\affiliation{Department of Physics and Center for
Theoretical Sciences, National Taiwan
University, Taipei 10617, Taiwan}
\affiliation{Leung Center for Cosmology and Particle Astrophysics\\
National Taiwan University, Taipei 106, Taiwan}
\author{Pisin Chen\footnote{chen@slac.stanford.edu}}
\affiliation{Department of Physics and Center for 
Theoretical Sciences, National Taiwan
University, Taipei 10617, Taiwan}
\affiliation{Leung Center for Cosmology and Particle Astrophysics\\
National Taiwan University, Taipei 106, Taiwan}
\affiliation{Kavli Institute for Particle Astrophysics and Cosmology\\
SLAC National Accelerator Laboratory, Menlo Park, CA 94025, U.S.A.}
\begin{abstract}
In this paper we study in detail the effect
of our recently proposed model of parity and charge-parity (PCP) violating 
varying alpha on the Cosmic Microwave Background (CMB) photon 
passing through the intra galaxy-cluster medium (ICM). 
The ICM is well known to be composed of magnetized plasma. 
According to our model, the polarization and intensity
of the CMB would be affected when traversing through the ICM
due to non-trivial scalar photon interactions.
We have calculated the evolution of such polarization and intensity
collectively, known as the stokes parameters of the CMB photon 
during its journey through the ICM and tested our results against 
the Sunyaev-Zel'dovich (SZ) measurement on Coma galaxy 
cluster. Our model contains a PCP
violating parameter, $\beta$, and a scale of alpha variation $\omega$.
Using the derived constrained on the photon-to-scalar conversion 
probability, ${\bar P}_{\gamma \rightarrow \phi}$, for 
Coma cluster in ref.\cite{shaw} we found a contour plot
in the ($\omega,\beta$) 
parameter plane. The $\beta =0$ line in this parameter space
corresponds to well-studied Maxwell-dilaton type models which
has lower bound on $\omega \gtrapprox 6.4 \times 10^{9}$ GeV. 
In general, as the absolute value of $\beta$ increases, 
lower bound on $\omega$ also increases. 
Our model in general predicts the modification of the 
CMB polarization with a non-trivial dependence
on the parity violating coupling parameter $\beta$. 
However, it is unconstrained in this particular study. We show that this effect can in principle be detected in the
future measurements on CMB polarization such that $\beta$
can also be constrained.

\end{abstract}

\maketitle
\section{Introduction}
There has been growing interests in the recent past 
to extend all the standard model of particle physics and test against
the present day high precession measurements. Parity violation
has already been proved to be one of its simplest and 
straightforward extension. 
It is already a well-established fact that there exists
parity (P) and charge-parity (CP)  violation in the electroweak sector. 
This particular observation drives people for the last several years
to study various different possible sources of PCP violation beyond
the standard model \cite{carroll,marc,contaldi,soda,debu}.
The basic idea of all these models is to add an explicit
parity violating term in the Lagrangian. Interestingly all those
different PCP violating models predicts different 
potentially observable phenomena such as cosmic
birefringence \cite{carroll,marc} and left-right asymmetry
in the gravitational wave dynamics \cite{contaldi,soda} which
could be detectable in the future experiments.
Recently we have also constructed a parity and
charge-parity (PCP) violating model \cite{debupisin}
in the framework of ``varying alpha theory" 
with the advantage over that of other scalar field model 
such as Carroll's in that the origin of the parity 
violation may be better physically motivated.

String theory gives us ample evidences to 
consider theories of varying fundamental constants in
nature. Since string theory is actually a higher dimensional
theory, all the fundamental constants
are emergent because of dimensional reduction.
So our hope is that future 
high precession cosmological as well as laboratory experiments may
provide some signatures of new physics, which also includes
the variation of fundamental constants.

After the proposal of a consistent 
framework of variation of fine structure constant $\alpha$ 
by Bekenstein \cite{bek2}, an extensive effort have been made 
for the last several years on the theoretical \cite{bsm,bsm1,barrow}
as well as the observational side 
\cite{murphy,webb,webb2,dent,landau} of this $\alpha$ variation.
The important point to mention that
it is not the other observable effect but the effect of direct
fine structure constant variation on the cosmology which has been 
considered extensively in the literature.
What we want to emphasize is that our recently proposed 
PCP violating extension to this model opens up the 
possibility to test it against various other observable 
effect apart from just the variation of $\alpha$ in 
the direct laboratory measurement \cite{debupisin2} as well
as indirect cosmological measurement. This is the indirect cosmological
measurement leading to the stringent constraint on a varying alpha model  
which is the main subject of study of our present paper.
We have already put constraints on our model parameters space against
various laboratory experiments like BFRT \cite{bfrt}, PVLAS \cite{pvlas} 
and Q\&A \cite{qa}.
The main goal of all these experiments is to measure the change of
states of a polarized laser beam propagating through 
the region of externally applied magnetic field. 
External magnetic field induced polarization in a model of 
scalar(pseudo scalar) coupled with electromagnetic field has been
the subject study for a long time 
\cite{raffelt,zavattini,pdpj,birch,kendall,cf,cfj,ralston,
dmssssg,ng,lepora,cowh,feng,maravin,shaw,schelpe}.
The model that we recently introduced also exhibits this effect
induced from the PCP violating term in our varying fine structure
constant theory \cite{debupisin}.
This motivates us to use a different class of experiments
to constrain the parameters of a given
varying fine structure constant theory. Such approach
has not been explored before. 

In terms of simple well know dilaton or axion electrodynamic models
our model can be thought of as a natural generalization of all those
where we have both parity even and parity odd coupling with photon.
More importantly it is not just an arbitrary addition but
a basic well known underlying assumption of varying fine structure 
constant which dictates to us the form of the 
scalar field coupling function with the electromagnetic field up to
some unknown constant which will be determined from the observation.
So, from our current study not only can we constrain
those constant parameters but also can shed some 
light on the possible variation of the fine structure constant
over a cosmological time. Our current study will be 
particularly focused on CMB observation and how it constrains 
our model parameter in the same spirit as of all the previous studies
separately on the scalar or axion electrodynamics models.
In this regard our study can, therefore, be thought 
of as a coherent study of all those scalar and axion 
electromagnetic models studied so far. Our model has two independent
parameters namely $\omega$ and $\beta$. It is the parameter 
$\beta$, the ratio between 
axion and scalar type coupling with photon field, which 
parametrizes the PCP violating coupling strength. In the present study,
we will see how this PCP violating parameter $\beta$ effects 
various observable quantities.  

In our previous work \cite{debupisin2} 
we put bounds on our model
parameters based on the 
birefringence and the dichroism of the vacuum induced from
the non-trivial coupling of a photon in the laboratory 
based experiment. We would like to point out that
the bounds we derived in our previous study is completely
excluded compared with the 
bound we found in the present study. This essentially
tells us that with the current experimental parameter values 
it is impossible to see any signal of birefringence and the dichroism in
those laboratory experiments. 

In this paper, we will be exploring another class of 
cosmological observations to constrain our model parameters.
We will analyze the effect of our PCP violating varying fine structure
constant model on the CMB photon when passing through the 
ICM. From various cosmological observations it has already been 
verified that ICM consists of strong magnetized plasma with
the magnetic field up to 30 $\mu$G. In the presence of this ICM plasma, the CMB
photons encounter an inverse Compton scattering
with the electron. This effect is known as SZ effect.
This scattering process does not affect the number density 
but changes the energy distribution of the
incoming CMB photons. As it is well known, CMB photons coming
from the last scattering surface encode a wealth of information
related to the properties of structure formation and more
importantly the information about the inflationary dynamics in the
very early universe.
All the important effects on CMB photon after the last
scattering, therefore, should be carefully investigated.
As we just mentioned, the SZ effect is one of those which have already been
studied quite intensively. If there exists some light scalar field
that couples to a photon, then we should be able to 
see the modification of the CMB spectrum 
due to the non-zero photon-to-scalar conversion 
probability amplitude in the 
presence of background magnetized plasma. There exist many
different models where this phenomena can occur. In this regard, 
standard axion-photon and dilaton-photon
system have been studied quite extensively from
theoretical as well as phenomenological
point of view \cite{dmssssg,ng,lepora,cowh,feng,maravin}.
Another model called chameleon model \cite{khoury} has also 
non-trivial effect on CMB \cite{shaw,schelpe}. In our model
which is the generalisation of the 
Bekenstein-Sandvik-Barrow-Magueijo 
(BSBM) theory of the varying
fine structure constant, has also natural coupling
between scalar and photon with parity violation. In this
paper we will explore in detail the effect of our model on 
the CMB photon passing through the ICM magnetized plasma, 
with emphasis on the effect of PCP violation. 

We organize this paper as follows: in Section \ref{sec1}, after 
briefly reviewing our PCP violating ``varying alpha theory" \cite{debupisin}, 
we will analyse in detail the optical properties and calculate the evolution
of the stokes parameter of the electromagnetic wave when it is passing
through the magnetized plasma.       
In the subsequent section \ref{sec2}, we will analyse 
the effects of our model on CMB photon. We calculate 
the evolution of stokes parameters
of CMB photon when passing though the ICM magnetized plasma. 
As we have mentioned before, we will test our result against
the SZ measurement of a particular galaxy cluster, the
Coma Galaxy cluster. The model
of the ICM magnetized plasma we will be using is the well-known
power spectrum model. We will analytically calculate
approximate expression for the stokes parameters of the incoming CMB photon
after passing through the ICM magnetic field and plasma of a
galaxy cluster. Then in section \ref{sec3}, after briefly reviewing
the general properties the galaxy cluster magnetic field, we 
will use our approximate expression of the 
photon-to-scalar conversion probability, ${\bar P}_{\gamma \rightarrow
\phi}$, which is responsible for the additional modification of 
the CMB temperature over the standard SZ effect, to
constrain our model parameter. 
We will use the derived upper bound on ${\bar P}_{\gamma \rightarrow
\phi}$ from the Coma cluster in the reference \cite{shaw} to constrain
the scale of variation of fine structure constant $\omega$.   
Subsequently in section \ref{sec4} we discuss about the modification
to the polarization stokes parameter of the CMB photon and its observational
aspects. Until now we do not have any observation on the change
of polarization of the CMB photon due to the ICM mainly because 
of experimental difficulty. We suggest that several 
recent experiments on the polarization measurement 
such as STP-Pole, ALMA, POLAR, which are either ongoing or 
under development, with there high angular resolution 
can in principle help to constrain 
the parameter space of our model. 
Concluding remarks and future prospects are provided in
Section \ref{con}.

\section{Optics in a PCP violating varying alpha theory} \label{sec1}
A varying alpha theory \cite{bek2,bsm,bsm1}
is usually referred to as a theory of spacetime variation of the 
electric charge of any matter field. 
The fine-structure constant in such a theory, therefore, conveniently
parametrazied by $\alpha = e_0^2 e^{2\phi(x)}$ in natural units.
According to above definition this theory enjoys a shift
symmetry in $\phi$ i.e. $\phi\rightarrow \phi + c$ and
also the modified U(1) gauge transformation
$e^\phi A_\mu \rightarrow e^\phi A_\mu +\chi_{,\mu }.$
So, an unique gauge-invariant, shift-symmetric 
and PCP violating Lagrangian for the modified scalar-electromagnetic fields
can be written as
\be
{\cal L} ~=~  M_p^2 R - \frac {\omega^2} 2 \pr_{\mu}\phi
\pr^{\mu} \phi ~ - \frac 1 4 e^{-2\phi }
F_{\mu\nu} F^{\mu\nu}
+ \frac {\beta}{4}e^{-2\phi } F_{\mu\nu} {\tilde F}^{\mu\nu}~+~
{\mathcal L}_m,    \label{action}
\ee
where electromagnetic field strength tensor can be expressed as
\bea
F_{\mu \nu }=(e^\phi {\bf a}_{\nu} )_{,\mu }-(e^\phi
{\bf a}_{\mu} )_{,\nu } = {\bf A}_{\nu,\mu }- {\bf A}_{\mu,\nu }.
\eea
with ${\bf A}_{\mu} = e^{\phi} {\bf a}_{\mu}$ as a modified
electromagnetic gauge potential.
$R$ is the curvature scalar and $\beta$ is
the PCP violating coupling parameter to be determined from the 
observation. we also set $e_0=1$ for convenience.
As can be easily seen, the above action reduces to the usual
form when $\phi$ is constant. The parameter $\omega$ sets a
characteristic scale of the theory above which one expects
Coulomb force law to be valid for a point charge.
Shift symmetry protects the scalar field not to have
any arbitrary potential  function in our Lagrangian. 
Of course one can break this shift symmetry by introducing a potential term, 
which has recently been studied in 
\cite{bali}. We will leave this 
for our future study in the context of PCP violating
varying alpha theory.

In this section we will do the general analysis in 
detail on the scalar-photon mixing 
phenomena in the background plasma with magnetic field. 
Our study would be 
relevant to the present day and also future various precision 
cosmological as well as astrophysical optical measurements.
The Maxwell and scalar field equations turn out to be of standard type 
with the modifications coming from non-trivial scalar field $\phi$ coupling .
\bea
&&\Box \phi = \frac {e^{-2 \phi}}
{2 \omega^2}\left[- F_{\mu \nu }F^{\mu \nu} ~+~ 
\beta F_{\mu \nu } \tilde{F}^{\mu \nu} \right],\\
&&\nabla \cdot {\bf E} = - (-2 \nabla { \phi} \cdot {\bf E} +
4 \beta \nabla { \phi} \cdot {\bf B} ), \nonumber \\
&&\partial_{\eta}({\bf E}) - \nabla \times {\bf B} = 2
 ({\dot { \phi}} {\bf E} -
\nabla { {\phi}}\times {\bf B}) -4 \beta( {\dot{{\phi}}}
{\bf B} + \nabla {{\phi}} \times {\bf E}),\nonumber \\
&&\bf \nabla \cdot \bf B = 0,\nonumber \\
&&\partial_{\eta}\bf B + \bf \nabla \times \bf E = 0,\nno \\
\eea
 
As is well known from various measurements, 
at cosmological as well as astrophysical scales there exists a
background magnetic field which may have a 
significant effect on the electromagnetic field coming from 
various sources. In this paper we are particularly interested
in studying the effect on the CMB photon. Studying 
the effect of some other external field
on the CMB photon is of particular interest because of its
prime importance in cosmology.With this motivation in mind, 
we will try to calculate the
effect of magnetized plasma background on 
the electromagnetic wave.  In terms of vector potential i.e. 
${\bf B}=\nabla \times {\bf A}$, the above equations  
can be written in the following suitable form  

\bea \label{equations}
&&(\nabla^2  - \pr_t^2 ) {\bf A} =- 4 \beta {\bf B} \pr_t \phi - 2 
(\nabla \phi \times {\bf B}) \nno \\
&&(\nabla^2 -\pr_t^2 ) \phi = \frac {2 {\bf B}^2}{\omega^2} \phi
- \frac {2 }{\omega^2} {\bf B} \cdot (\nabla \times {\bf A}) + 
+\frac {4 \beta }{\omega^2} {\bf B} \cdot \pr_t {\bf A} 
\eea
In this case we assume the background magnetic field is ${\bf B}$.
Because of smallness of the effect we consider linear order equations
for the scalar-photon system. 
In the above derivation we use the gauge condition $\nabla \cdot {\bf A} = 0$
and consider the scalar potential ${\bf A}_0 = 0$. 
Now, assuming the propagation direction of the electromagnetic wave 
to be in the z direction, we
take the form the fields ansatz to be
\bea
{\bf A}(z,t) = {\bf A}^0 e^{- i \varpi t }~~~ ;~~~
 \phi(z,t)=\phi^0 e^{- i \varpi t}  
\eea
where ${\bf A}= \{{\bf A}_x,{\bf A}_y,0\}$. 
In order to solve them 
analytically, we will follow the same procedure as in \cite{raffelt}. 
We further assume that the background
magnetic field variation is very small compared to the scalar and
the photon frequency ${\varpi}$. With this assumption we can approximate
the dispersion operator to be 
\bea
\pr_z^2  + \varpi^2= (\varpi + i \pr_z)(\varpi - i \pr_z) = 
(\varpi+k) (\varpi + i \pr_z) \simeq 2 \varpi (\varpi + i \pr_z) ,
\eea
assuming the dispersion relation to be $k = n \varpi$
with $|n-1|\ll 1$. Therefore, we can write down the 
above system of eqs.\ref{equations} as
\bea
&&(i \pr_z  + \varpi) {\bf A}_x -  i ({\bf B}_y + 2\beta {\bf B}_x) \phi = 0 \\
&&(i \pr_z + \varpi) {\bf A}_y + i ({\bf B}_x - 2 \beta {\bf B}_y) \phi==0 \\
&&(i \pr_z + \varpi)\phi - \frac { {\bf B}^2}{\omega^2 \varpi} \phi
- \frac i {\omega^2} ({\bf B}_x - 2 \beta {\bf B}_y) {\bf A}_y + 
\frac i {\omega^2} ({\bf B}_y + 2 \beta {\bf B}_x) {\bf A}_x =0
\eea
Now if we take into account the plasma effects, the above
set of linear equations can be expressed as follows:  
\bea \label{eqo}
&&( i \frac d {dz} + {\cal M} )\left(\begin{array}{c}
 {\bf A}_x\\ {{\bf A}}_y\\ {\Phi}\\
\end{array}\right)=0\\
\mbox{Where}~~~~~&& {\cal M} =\begin{bmatrix}  \varpi +
\Delta_x &0 &  - i ({\bf B}_y + 2\beta {\bf B}_x) \\
0&  (\varpi+\Delta_y) & i ({\bf B}_x - 2 \beta {\bf B}_y)  \\
\frac i {\omega^2} ({\bf B}_y + 2 \beta {\bf B}_x)  &
- \frac i {\omega^2} ({\bf B}_x - 2 \beta {\bf B}_y)  
& \varpi - \frac { {\bf B}^2}{\omega^2 \varpi}\\
\end{bmatrix} \nno.
\eea 
Here ${\cal M}$ is called scalar and photon mixing matrix. The new
notation are
\bea
\Delta_{x,y}=
\Delta_{QED} + \Delta_{CM} + \Delta_{plasma}
\eea
All the terms in the second expression of the above equation have been
considered before in the axion-photon study. $\Delta_{QED}$ comes
from the effect of vacuum polarization giving rise to the refractive index
of photon. This term is also known to be associated 
with the lowest order non-linear Maxwell Lagrangian 
(Euler-Heisenberg term). $\Delta_{CM}$ is known as Cotton-Mouton term
which is the effect of birefringence of gases and liquids in the 
presence of a magnetic field. The last term is
due to the background plasma through which the photon traversed.
The usual expression for those terms are as follows:
\bea
&&\Delta_{QED}^x = \frac 7 2 \varpi \zeta, ~~~\Delta_{QED}^y 
= 2 \varpi \zeta \nno\\
&&\Delta_{CM}^x-\Delta_{CM}^y = 2 \pi C {\bf B}_0^2, ~~~\Delta_{plasma}=
-\frac {\varpi_{plasma}^2}{2 \varpi} = 4 \alpha_0 \frac {\rho_e}{m_e} \frac 1 {2
\varpi},
\eea
where
$\zeta =  (\alpha_0/(45 \pi) ({\bf B}_0/{\bf B}_c)^2, {\bf B}_c 
\equiv m_e^2/e =
4.41 \times 10^{13}$ G the critical field strength,
 $m_e$ the electron mass, $e$ the electron
charge and $\alpha$ the fine structure constant. Note that in the 
above expression for $\zeta$, we ignore the correction due to
the fine structure constant variation as it contributes to the
higher order in fluctuation in Eq.\ref{equations}.   

Now in order to solve the above set of equations
we define
\bea
{\bf A} = e^{- \varpi(t-z) + i \zeta(z)} {\bar{\bf A}}~~~;~~~
\phi  = e^{- \varpi(t-z) + i \zeta(z)} {\bar \phi}
\eea
where $\zeta'(z) = -{\varpi^2_{plasma}}/{2 \varpi}$.
With these new variables the above set of equations can be written as follows

\bea \label{eqs}
&&( \frac d {dz} +  {\bar M} )\left(\begin{array}{c}
 \bar{{\bf A}}_x\\ \bar{{\bf A}}_y\\ e^{i S} {\Phi}\\
\end{array}\right)=0\\
\mbox{Where}~~~~~&& {\bar M} =\begin{bmatrix}  
0 &0 &  - ({\bf B}_y + 2\beta {\bf B}_x) e^{-i S} \\
0& 0 &  ({\bf B}_x - 2 \beta {\bf B}_y) e^{-i S}  \\
\frac 1 {\omega^2} ({\bf B}_y + 2 \beta {\bf B}_x) e^{i S} &
- \frac 1 {\omega^2} ({\bf B}_x - 2 \beta {\bf B}_y) e^{i S}
& 0\\
\end{bmatrix} \nno
\eea
Where we define
\bea
S(z) = -\int_0^z \left( \frac {\varpi^2_{plasma}}{2 \varpi} -
 \frac { {\bf B}^2}{\omega^2 \varpi} \right) dx
\eea 
In order to solve the above equations of motion we will
make an approximation following Ref.\cite{shaw}, where the amplitude
of the mixing matrix is small, i.e, $Tr[{\cal M}{\cal M}^{\dagger}] <1$.
Let us consider the solution of the form
\bea
\left(\begin{array}{c}
 \bar{{\bf A}}_x(z)\\ \bar{{\bf A}}_y(z)\\ e^{i S(z)} {\Phi}(z)\\
\end{array}\right)  \simeq ({\mathbb I} + {\mathcal J}_1 + 
{\mathcal J}_2 +\cdots) \left(\begin{array}{c}
 \bar{{\bf A}}_x(0)\\ \bar{{\bf A}}_y(0)\\ e^{i S(0)} {\Phi}(0)\\
\end{array}\right)
\eea
where 
\bea 
{\mathcal J}_1= \int_0^z {\bar M(x)} dx =\begin{bmatrix}
0 &0 &  - {\cal B}_y^* \\
0& 0 &  {\cal B}_x^*  \\
\frac {{\cal B}_y} {\omega^2}  &
- \frac {{\cal B}_x} {\omega^2}
& 0\\
\end{bmatrix} ~~~;~~~
{\mathcal J}_2= \int_0^z {\bar M}'(x) {\bar M(x)} dx
\eea
where
\bea
{\cal B}_i = \int_0^z ({\bf B}_i -2 \beta \epsilon^{ij} {\bf B}_j) e^{i S} dx
\eea
with $\epsilon^{xy}=1, \epsilon^{yx}=-1$.
Once we know the approximate solution, we can write down the polarization
states of the electromagnetic field under study in terms of Stokes parameters
\bea
I(z) &=& |{\bf A}_x|^2 + |{\bf A}_y|^2 \\
Q(z) &=& |{\bf A}_x|^2 - |{\bf A}_y|^2 \\
U(z) &=&  2 Re({\bf A}_x^*{\bf A}_y) \\
V(z) &=&  2 Im({\bf A}_x^*{\bf A}_y) 
\eea
where, $I$ is intensity, $Q(z), U(z)$ are linear polarization and
$V(z)$ is circular polarization of the electromagnetic field.
After traversing the path length z, polarization states take the 
following explicit form
\bea
I(z) &=& I(0)(1 - P_{\gamma\rightarrow \phi}) + Q(0) {\cal Q}(z) 
+U(0) {\cal U}(z) - V(0) {\cal V}(z),\\
Q(z) &=& Q(0)(1 -P_{\gamma\rightarrow \phi}) + I(0) {\cal Q}(z) 
+ U(0) ({\cal U}(z)-2 {\cal L}_1(z))  - V(0) ({\cal V}(z) -2 {\cal L}_2(z)),
\nno\\
U(z) &=& U(0)(1 -P_{\gamma\rightarrow \phi}) + I(0) {\cal U}(z)
- V(0) (P_{\gamma\rightarrow \phi}-2 {\cal L}_3(z))  
- Q(0) ({\cal V}(z) -2 {\cal L}_1(z)), \nno \\
V(z) &=& V(0)(1 -P_{\gamma\rightarrow \phi}) + I(0) {\cal V}(z)
- U(0) (P_{\gamma\rightarrow \phi}-2 {\cal L}_3(z)) 
- V(0) ({\cal V}(z) -2 {\cal L}_2(z)), \nno
\eea
where we have defined 
\bea
P_{\gamma\rightarrow \phi} &=& \frac {1}{2 \omega^2} (|{\cal B}_x|^2 + |{\cal B}_y|^2)~~;~~
{\cal Q}(z) = \frac {1}{2 \omega^2} (|{\cal B}_x|^2 - |{\cal B}_y|^2),\\
{\cal U}(z) &=& \frac {1}{2 \omega^2} ({\cal B}^*_x {\cal B}_y + {\cal B}^*_y {\cal B}_x)~~;~~
{\cal V}(z) = \frac {1}{2 \omega^2} ({\cal B}^*_x {\cal B}_y - 
{\cal B}^*_y {\cal B}_x), \nno \\
{\cal L}_1(z) &=& 
\frac {1}{2 \omega^2}\int_0^z ({\cal B}^{*'}_x 
{\cal B}_y + {\cal B}'_x {\cal B}^*_y)~~;~~
{\cal L}_2(z) = \frac {1}{2 \omega^2}\int_0^z 
({\cal B}^{*'}_x {\cal B}_y - {\cal B}'_x {\cal B}^*_y), \nno \\
{\cal L}_3(z) &=& 
\frac {1}{2 \omega^2}\int_0^z ({\cal B}^{*'}_x {\cal B}_x + 
{\cal B}'_y {\cal B}^*_y), \nno
\eea
In the above expressions we assume the initial correlations as follows:
\bea
<\phi^*(0){\bf A}_i(0)> =0 ~~;~~ <{\bf A}^*_i(0) \phi(0)> =0~~;~~
<\phi^*(0)\phi(0)> =0 
\eea
The photon-to-scalar or scalar-to-photon transition amplitude is defined
by $P_{\gamma\rightarrow \phi} (z)$. 
Variation of fine structure constant leads to an effective change
in photon intensity, which in turn effects on the CMB temperature.
It can also induce polarization of the photon when 
traverses a long cosmological distance. We have already 
discussed in the introduction that
we are interested to study the effect of 
ICM on the CMB photon traversing
through it. Therefore, in order to make an estimate of the amount of 
effect due to the variation of fine structure constant, we have to
consider a specific model of magnetic field ${\bf B}$ variation and also the 
electron density $\rho_e$ variation at the galaxy cluster scale. As the CMB
photon passes through the ICM, its frequency 
distribution changes due to inverse Compton scattering 
SZ effect with the electrons in the plasma. 
This essentially means the non-vanishing
photon to scalar  conversion probability $P_{\gamma\rightarrow \phi} (z)$.  
In this paper we will estimate the effect due to this conversion
probability considering particular model of magnetic field 
and electron density variation in the galactic medium closely 
following the reference \cite{shaw}.
The particular model that we will be considering is Power spectrum model
for the spatial variation of galaxy cluster magnetic field ${\bf B}$ 
and the electron density $\rho_e$. We will also be discussing
about the effect on the polarization states of the CMB photon.
It is well known that the initial states of the CMB photon are
very lightly polarised compared to its intensity. According to
the observation fractional linear polarization compared to the
intensity parametrized by $\langle Q(0)^2 \rangle^{1/2} /I(0),
\langle U(0)^2 \rangle^{1/2} /I(0) \sim {\cal O}(10^{-6})$ and 
the fractional circular polarization 
$\langle V(0)^2 \rangle^{1/2} /I(0) \ll {\cal O}(10^{-6})$.
So essentially the change of states of the CMB photon after traversing a 
long intergalactic distance $z$ is proportion to the initial 
intensity $I(0)$ and conversion probability 
$P_{\gamma\rightarrow \phi} (z)$. 
\bea \label{pquv}
I(z) &\simeq& I(0)(1 - P_{\gamma\rightarrow \phi}) ~~~;~~~
Q(z) \simeq I(0) {\cal Q}(z) \nno \\
U(z) &\simeq& I(0) {\cal U}(z) ~~~;~~~V(z) \simeq  I(0) {\cal V}(z)
\eea
Once we get the above approximate expression for the
stokes parameters for the electromagnetic wave, we can 
analyse the effect on CMB photon which is believed to be one of
the important probes to understand the cosmology. As we 
have discussed before, the framework that we built up is 
readily applicable to analyse the evolution
of states of CMB photon passing
through the galaxy clusters. In the following sections 
we will apply our framework and do the quantitative estimates 
of the temperature as well as polarization modulations on 
the CMB photon due to ICM magnetized plasma fields.

\section{The power spectrum model and the effect on CMB} \label{sec2}
The most realistic model for the magnetic field ${\bf B}$ 
and the electron density $\rho_e$ in a galaxy cluster 
is described by the so-called power spectrum model \cite{vogt1}. 
The most relevant physical quantities are the two point correlation 
functions of $\delta {\bf B}_i$ and $\delta \rho_e$, defined by
\bea
R_{{\bf B}ij}(x) &=& <\delta {\bf B}_i(y)\delta {\bf B}_i(x+y)> = 
\frac {1}{4 \pi} \int d^3 k P_{{\bf B}ij}(k) e^{i {\bf k}\cdot {\bf x}},\\
R_{{\bf e}}(x) &=& <\delta \rho_e(y)\delta \rho_e(x+y)> = 
\frac {1}{4 \pi} \int d^3 k P_{e}(k) e^{i {\bf k}\cdot {\bf x}} .
\eea
In the power spectrum model the magnetic field fluctuation component is
approximately assumed to be gaussian random variable, i.e 
$<\delta {\bf B}_i> = 0$, where the average is taken over full spatial
length of a galaxy cluster through which the CMB photon is propagating.
With this assumptions one can show $P_{{\bf B}ij}(k) = 
\frac 1 3 \delta_{ij} P_{\bf B}(k)$. In the above expression for the
power spectrum we also assumed that the fluctuations are approximately 
position independent.
The corresponding correlation lengths for the fluctuation of electron density
and the magnetic field are defined by \cite{murgia}
\bea
L_{\bf B} = \frac {\int_0^{\infty} k dk P_{\bf B}(k)}
{2\int_0^{\infty} k^2 dk P_{\bf B}(k)}~~~;~~~L_{e} = \frac 
{\int_0^{\infty} k dk P_{e}(k)} {2\int_0^{\infty} k^2 dk P_{e}(k)}
\eea
Now in order to estimate the modified polarization of states and intensity
of a CMB photon after traversing a distance $L$ through the
galaxy cluster, the basic quantity we have to calculate is
$G_{ij}(x) = < {\cal B}^*_i(y) {\cal B}_j(y+x)>$. The main goal is to 
express the above correlation functions in terms of two power spectra
$P_{\bf B}(k)$ and $P_e(x)$. As we mentioned before, in order to calculate
this we will closely follow the procedure of \cite{shaw}.

In order to avoid complications in our main text, we only 
quote our essential expressions which are directly related to the observable
quantity. All the detail calculations have been given in the appendix.
One can see that to the leading order in $1/{\omega^2}$ and $\beta$, 
the photon-to-scalar conversion probability 
can be written in the following compact form
\bea 
{\bar P}_{\gamma\rightarrow \phi} = {\bar P}^{reg}_{\gamma\rightarrow \phi}+ 
{\bar P}^{ran}_{\gamma\rightarrow \phi} +
\frac {8\beta^2} 3 \left(\int_{{\bar \Delta}}^{\infty} +
 \int_{{\bar \Delta}'}^{\infty}\right) k dk {\cal F}^k_{(1)}.
 \nno 
\eea
To simplify our further calculations in the above 
expression we consider the propagating photon with
a single frequency $\varpi$ so that ${\bar{\Delta}} = {\bar{\Delta}}'$. .
For convenience, we have separated the total scalar conversion 
probability amplitude into a term coming from the regular("reg") 
ICM magnetic field ${\bf B}_0$
and the other terms coming from the random("ran") 
magnetic field $\delta {\bf B}$. We will provide the expressions 
for ${\bar P}^{reg}_{\gamma\rightarrow \phi}$ and ${\bar P}^{ran}_{\gamma\rightarrow \phi}$ 
in our subsequent discussions.  
With these new definitions on the part of scalar to photon transition probability
one can also check that the expression for the induced polarization of the
CMB photon, after traveling through the ICM of length $L$, becomes
\bea
{\bar V}(L) &\simeq& -\beta I(0) {\bar P}^{ran}_{\gamma\rightarrow \phi},\\ 
{\bar Q}(L) &\simeq& I(0)  {\bar P}^{reg}_{\gamma\rightarrow \phi}
 (\cos 2 \theta - 
4 \beta \sin 2 \theta), \nno \\
{\bar  U}(L) & \simeq & I(0) {\bar P}^{reg}_{\gamma\rightarrow \phi} 
 (\sin 2 \theta +   
4 \beta \cos 2 \theta).
\eea
One can immediately see that in addition to the 
standard scalar-photon coupling contribution, all the observable quantities 
depend non-trivially on the PCP violating parameter $\beta$.

In order to calculate the amount of effect of the 
magnetized plasma on the incoming CMB photon,
we need to consider observed power spectrum $P_{\bf B}(k)$ and 
$P_e(k)$ of the magnetic field and
electron density respectively. On the small scales it
is customary to parameterize the power spectrum by 
power law in momentum space as follows:
\bea \label{ps}
k^2 P_{\bf B}(k) = {\cal P}_{\bf B} \left (\frac k {k_0}\right)^{\gamma}~~~;~~~
k^2 P_{e}(k) = {\cal P}_{e}^2  k^{\gamma},
\eea
where ${\cal P}_{\bf B} $ and  $P_{e}(k) $ are the normalization 
constants. $\gamma < -1$ and $k_0 = 1 kpc^{-1}$. 
A special universal value $\gamma = -5/3$ on 
small scale corresponds to the well-known spectral index 
for the three dimensional Kolmogorov's theory of turbulence.
We also assume that this power law form holds for a wide range
of scales of the magnetic and electron density fluctuations in
the ICM. Interestingly, observations on
many different galaxy clusters suggest that on a wide range of spatial
scales, the power spectrum is consistent with the Kolomogorov one. 
It is, therefore, straight forward to calculate
\bea
\int_{{\bar \Delta}}^{\infty}  k dk P_{\bf B}(k) \propto 
\int_{{\bar \Delta}}^{\infty}  k dk P_{e}(k) \propto {{\bar \Delta}}^{\gamma},
\eea
where ${\bar \Delta}  = \left(\frac {\varpi_{plasma}^2}
{2  \varpi} - \frac
{{\bar {\bf B}_0}^2} {\varpi \omega^2}\right)$. The 
critical length scale ${\bar \Delta}$ composed of
two part, the first part which is related to the
well-known quantity called plasma frequency, 
is dependent on the average electron density,
and the other part is dependent on the background magnetic field strength.
The latter part also depends on the scale of fine structure
constant variation $\omega^2$. In terms of the length scale, we can write
\bea \label{estimate1}
\frac {\varpi_{plasma}^2} {2  \varpi} &\simeq& 0.208\times 10^2 \frac 1 {pc}
\left(\frac {2\pi 100~ \mbox{GHz}}{\varpi}\right)
\left( \frac {{\bar \rho}_e}{10^{-3} \mbox{cm}^{-3}}\right), \nno \\
 \frac {{\bar {\bf B}_0}^2} {\varpi \omega^2} &\simeq&
4.29 \times 10^{-6} \frac 1 {pc} \left(\frac {2 \pi 100~ \mbox{GHz}}
{\varpi}\right) \left(\frac {{\bar{\bf B}}_0}{30 \mu \mbox{G}}\right)
\left(\frac {1~ \mbox{GeV}}{\omega}\right)^2 .
\eea 
It can be easily observed from the above expressions that the 
value of magnetic field dependent part is in fact very small, 
even for $\omega \simeq {\cal O}(1)$ GeV, compared to the
plasma frequency part. This is also in accord with our previous
perturbative expansions of the various magnetic correlation functions
in terms of standard two point correlation function.
As is known for a typical
galaxy cluster, if we consider the CMB photon frequency, 
$\omega \approx 30-300$ GHz then
inverse of the first line of the Eq.\ref{estimate1} takes the 
approximate value $\simeq 10^{-3}-0.1$ pc.
Therefore, all the observable quantities like $P_{\gamma \rightarrow \phi}, 
Q, U$ and $V$, which are sensitive to the 
critical length scale, is controlled by the plasma frequency of the
Intra-galactic plasma \footnote{1 pc$ = 3.08568025\times
10^{18} cm$; 1 $cm^{-1} = 1.98\times 10^{-14}$ GeV; 1 $\mu$G $=10^{-6}$ G 
$= 1.95 \times 10^{-26}$ GeV$^2$ }. It is important point to note that
the measurement, so far, probes the power spectrum at the 
spatial scales larger than the few kiloparsecs. 
In order to proceed further we will assume, therefore, 
that the power spectrum Eq.\ref{ps},
which  holds for the momentum $k>k_{*}$, also includes the critical
length scale ${\bar \Delta}$. 
On the other hand for $k<k_{*}$, the power spectrum
remains almost constant in consistent with the observation \cite{vogt}. With
these assumptions, following \cite{shaw},
the normalization constant for the assumed power
spectrum ${\cal P}_{\bf B}$ and ${\cal P}_{e}^2$ can be approximately
estimated as        
\bea
k_0^{-\gamma} {\cal P}_{\bf B} \simeq \frac {2 
\left(\frac {\gamma}{\gamma+1} - \frac 1 x \right)^{\gamma} }
{\left(\log(x) - \frac 1 {\gamma}\right)^{\gamma +1}} L_{\bf B}^{\gamma +1}
\langle \delta {\bf B}\cdot \delta {\bf B}\rangle, \\
{\cal P}_{e}^2 \simeq \frac {2 
\left(\frac {\gamma}{\gamma+1} - \frac 1 x \right)^{\gamma} }
{\left(\log(x) - \frac 1 {\gamma}\right)^{\gamma +1}} L_{e}^{\gamma +1}
\langle \delta \rho_{e} \delta \rho_{e}\rangle ,
\eea
where for the typical galaxies, the value of $x\sim 10-200$.
From the assumed log-normal distribution of the electron 
density, it is straightforward to check that $I_e = 1 + \langle
\delta \rho_e \delta \rho_e \rangle/{\bar \rho}^2_e$. For example, 
the electron density fluctuation measurement on our
own galaxy suggests the approximate value of $I_e \sim 1 -2$. 
In the subsequent analysis, we also assume 
$L_{\bf B} \approx L_e$ which is in accord with the 
various observations in different galaxy clusters. 

Our analysis so far revealed that the fluctuating component of the
ICM magnetic field $\delta {\bf B}$ and plasma 
density $\delta \rho_e$ fields play the crucial role in the modification 
of the CMB intensity and polarization
tensor. On the other hand, the background regular components 
$({\bf B}_0, {\bar \rho}_e$ of those
quantities set the inherent critical scale of the system. Using the 
above results, in the subsequent sections, we will estimate 
the amount of effects such as temperature variation (SZ-like effect) and 
also the induced polarization of the CMB making use of all the
known parameters for the Coma cluster.  


\section{Galaxy Cluster's Magnetic field and SZ-like
Effect} \label{sec3}
   Because of the non-trivial scalar-photon interaction which 
depends on the nature of the magnetic field and plasma distribution, 
we have seen that the energy spectrum as well
as the polarization states of the CMB photon change non-trivially
when traversing through ICM. Furthermore, in order to understand the early 
universe physics and also the physics of structure 
formation, it is essential to observe correct
initial state of the CMB photon at the last scattering surface(LSS). 
With that in mind, all the intermediate effects on CMB photon,
 from the LSS to the observer on Earth, 
should be taken into consideration. We have already seen in our model
that, because of 
the photon-to-scalar conversion probability, 
the modification of the intensity as well as polarization states 
of the CMB photon depends strongly on the strength and
distribution of the magnetic field and plasma density in ICM. 

It is an experimental fact that the ICM carries magnetic field
with strength as high as 30 $\mu$G \cite{taylor}. One of the
common methods to determine the magnetic field profile of the 
ICM is to the observation of the Faraday rotation of 
the plane of polarization of the electromagnetic
wave coming from the extended polarized radio sources either
behind or embedded within the galaxy cluster under study. 
Observations and numerical simulations suggests that
ICM magnetic field can be best described by the power spectrum 
model \cite{murgia} that we have considered in our above analysis. 
The basic underlying assumption behind this
power spectrum model is the statistical homogeneity and isotropy 
of the fluctuations of magnetic field and electron density 
on a large volume of the galaxy cluster under consideration.
The linearly polarized radio emission experiences a 
rotation of the plane of polarization when it traverses 
through the ICM with a background
magnetic filed which has a component along the line of propagation.
The observed polarization angle is proportional to the square of 
the wavelength and a quantity called RM, which is a proportionality 
constant. The mathematical
expression for RM along the line-of-sight in 
the $\hat{{\bf z}}$ direction of a source located at $z_s$ is 
\bea
RM(z_s) = a_0 \int_0^{z_s} \rho_e(x\hat{{\bf z}}) 
{\bf B}_z(x\hat{{\bf z}} ) dz ,
\eea
where $a_0 = \alpha_0^3/\pi^{1/2} m_e^2, {\bf B}_z$ is the magnetic field 
along the line-of-sight, and the observer position is at $x =0$.
  
In order to understand better about the magnetic field structure 
in the ICM, one needs to understand electron distribution as well.
It is observed that the shape of the electron density distribution
also has some correlation with the background magnetic
field in the medium.
Experimentally, for example from the ROSAT full-sky survey, 
the electron density distribution has been determined
from the X-ray surface brightness profile of the 
hot and diffused gas that fills
the ICM. It is well-known that the radial profile of electron density 
from the galaxy core could be well fitted to a $\beta$ profile
\cite{govoni}:
\bea
\rho_e(r) = \rho_0 \left(1+ \frac {r^2}{r_c^2}\right)^{-3 \beta/2}
\eea
where $\beta \sim {\cal O}(1)$ and positive. $\rho_0$ is the mean 
electron density and $r_c$ is the core radius of the galaxy cluster. 
For a typical galaxy cluster, the values of those parameters are
$r_c \sim 100-200 kpc, \beta \sim 2/3$, and $\rho_0 \sim 0.0001-0.01 ~ 
cm^{-3}$. In addition to the main component
of the electron density there exists a fluctuating component that
can be best characterized by using the power spectrum model. 
Standard magneto-hydrodynamic(MHD) simulation of galaxy cluster formation 
suggests that the total magnetic energy should follow the power law
behavior of electron density: $\langle{\bf B}^2 \rangle
 \propto \langle \rho_e \rangle^{\eta}$.
Various theoretical arguments and the observations \cite{dolag}
predict that $\eta \simeq 1$. 

As we have already mentioned before, the CMB
photon encounters an inverse Compton scattering
with the electrons in the ICM plasma. This effect is known as SZ effect.
This scattering process changes the energy distribution of the
incoming CMB photon. The CMB is believed to be
one of the important cosmological probes. 
So, all the important effects on CMB photon after the last
scattering should be carefully investigated. 
We just stated that SZ effect is one of those effects
which have already been
studied quite intensively. We have calculated in the previous section
that because of non-trivial scalar-photon interaction,
the frequency spectrum of the CMB changes
because of the non-zero conversion probability amplitude. 
There exist many different kind of scalar field 
models where this phenomena exists due to the 
non-trivial scalar-photon coupling. All those models 
have been studied quite extensively from 
the theoretical as well as phenomenological
point of view \cite{dmssssg,ng,lepora,cowh,feng,maravin}. 
All those fields are collectively known as axion-like particles (ALPs).
One of the interesting examples, which has recently been gotten
much attention, is known as chameleon field \cite{khoury}.
The effective mass of the chameleon depends on the 
density of the surrounding matter distribution. Therefore,
in the low density region of space, this chameleon field also 
plays like a ALP. Extensive studies have been done
on its effect on the cosmology, more specifically in the 
context of the present paper see \cite{shaw,schelpe}. As we
have stated before, our model of PCP violating varying alpha 
predicts non-trivial effect on the photon field traversing
through the magnetized plasma. So, we will be able to see 
how our parity violating coupling can lead to the various effects on 
the CMB photon. As we know the intensity of the CMB photon 
is related to its temperature. The variation of temperature, 
therefore, is related to the variation of 
the intensity as follows:
\bea \label{sz}
\frac {\delta T} {T_0} = \frac {(1 - e^{-\mu \varpi})}
{\mu \varpi} \frac {\delta I} {I_0},
\eea
where, the Boltzmann factor $\mu =  \frac 1 {k_B T_0}$ with
average CMB temperature $T_0\simeq 2.75 K$. We have seen before 
due to the scalar-photon coupling, the intensity of the CMB
photon changes due to the ICM magnetic field and 
the electron density distribution. The point we would like to emphasize that,
the variation of fine structure constant can lead to a new kind
of non-trivial foreground effect on CMB photon. 
The temperature variation of CMB due
to the scalar-photon conversion probability, can then be expressed as 
\bea
\frac {\delta T} {T_0} \approx \frac {(e^{-\mu \varpi}-1)}
{\mu \varpi} {\bar P}_{\gamma\rightarrow \phi}(L).
\eea
In terms of physical quantities, the expression for  
${\bar P}_{\gamma\rightarrow \phi}(L)$ turns out to be 
\bea
{\bar P}_{\gamma\rightarrow \phi}(L) =  
\left( \frac {I_e^3 {\bf B}_{*}^2 }{{\cal N}^2 \omega^2} -
\frac {{\bf B}_0^2 \cos(\bar{\Delta} L_{eff})}{ {\cal N}^2 \omega^2}\right)  
\varpi^2 \bar{\rho}_e^{-2} + \frac {\pi L_{eff} I_e^2 (1+4\beta^2)
{\cal N}^{\gamma}}
{2 \gamma \omega^2}\left( \frac {{\cal P}_e^2 {\bf B}_{*}^2}{{\bar \rho}_{e}^2}
+ \frac {2 I_e {\cal P}_{\bf B}}{3 k_0^{\gamma}}\right) 
\varpi^{-\gamma} \bar{\rho}_e^{\gamma}, \nno \\
\eea
where we define 
\bea 
{\bf B}_{*}^2 = {\bf B}_0^2 + 
\frac 2  3 \langle \delta {\bf B}\cdot \delta {\bf B} \rangle~~~;~~~
\bar {\Delta} = \frac {\bar {\rho_0} {\cal N}}{\varpi}.
\eea
For the power spectrum model typically $-2 \leq \gamma < -1$, which obviously
includes Kolmogorov model of three dimensional turbulence where the
exponent  $\gamma = -5/3$. Several observations suggest that 
the regular component, ${\bf B}_0$ of ICM magnetic field, 
is much smaller than that of the random part $ \delta {\bf B}$. For example 
the ICM magnetic field in the central region of the Coma 
cluster has been determined from the Faraday RMs measurement \cite{feretti}. 
The strength of the regular part of the magnetic field is 
estimated to be $0.2 \pm 0.1 \mu$G with the coherence length of the 
order of 200 kpc. Where as the 
strength of the random part is $8.5\pm 1.5 \mu$G with the coherence
length on much shorter scales of $L_{\bf B} \sim 1$kpc. If we assume that 
the coherence length, $L_e$, and the power spectrum of the electron
density fluctuation $\delta \rho_e/{\bar \rho}_e$ are proportional
to that of the magnetic fluctuation, then the 
approximate expression for the dominant contribution to the 
scalar-to-photon conversion probability becomes
\bea \label{sztemp}
{\bar P}_{\gamma\rightarrow \phi}(L)\approx 
\frac {2 I_e^3 \langle\delta {\bf B}\cdot \delta {\bf B}\rangle }
{3{\cal N}^2 \omega^2} \varpi^2 \bar{\rho}_e^{-2} + 
\frac {2\pi L_{eff} I_e^2(1+4\beta^2) {\cal N}^{\gamma}}
{6 \gamma \omega^2} \frac {{\cal P}_e^2 \langle\delta 
{\bf B}\cdot \delta {\bf B}\rangle }{{\bar \rho}_{e}^2}
\varpi^{-\gamma} \bar{\rho}_e^{\gamma}. \nno \\
\eea

At this point it is important to mention the behavior of 
the standard thermal SZ effect which changes the intensity of the CMB photon
in the following way \cite{birkinshaw}
\bea
\frac {\delta T} {T_0} = \frac {\kappa_B T_e}{m_e} \tau_0 \left(\mu \varpi 
\coth \left(\frac {\mu \varpi}2\right) -4\right),
\eea
with $T_e$ being the temperature of the electron in the ICM plasma. 
The quantity $\tau_0 =  \int \sigma_T \rho_e(z) dz$ is known as optical depth. 
$\sigma_T $ is called Thompson cross-section. One can see, therefore,
that the thermal SZ effect is linear in electron density $\rho_e$ compared
to the power law behavior of SZ-like effect due to 
the varying alpha scalar field. Power law type frequency dependence 
in eq.(\ref{sztemp}) compared to the non-power law type 
frequency dependence in the standard thermal SZ effect could in
principle be detectable from the observation in the future 
high precession experiments.

\begin{figure}[t!]
\includegraphics[width=3.80in,height=3.50in]{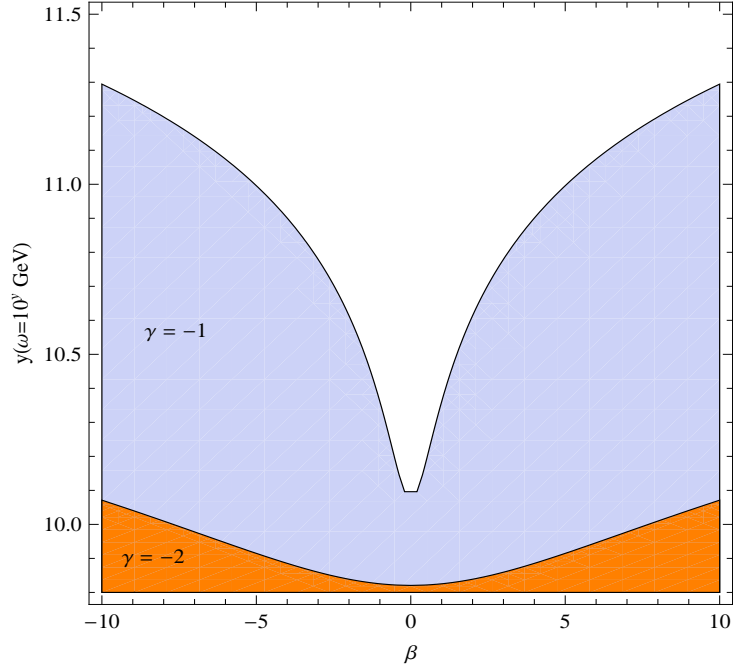}
\caption{\label{fig1} Bounds on PCP violating parameter
$\beta$ and the scale of varying fine structure constant $\omega$
using the possible bound on scalar to photon conversion probability 
$P^{coma}_{\gamma\rightarrow \phi}(204 \mbox{GeV}) < 6.2 \times 10^{-5}$ 
for the well known Coma galaxy cluster. 
This bound has been derived in \cite{shaw} from the SZ measurement 
on the Coma galaxy cluster.}
\end{figure}

In this paper, to constrain our model parameters, 
we will use the bound on the photon-to-scalar
conversion probability derived in the reference \cite{shaw}. 
The specific result that we are 
going to use is for the nearby Coma galaxy cluster. There exits
detailed measurements of the ICM magnetic field and also the SZ effect of 
this particular cluster by various experiments like OVRO, WMAP and MITO.
As has been derived in the reference \cite{shaw}, 
the upper bound on the photon-to-scalar conversion probability
for the Coma cluster is
\bea \label{szcoma}
P^{coma}_{\gamma \rightarrow \phi}(204 GHz) < 6.2 \times 10^{-5}
\eea
with the $95\%$ confidence level. It is also argued that the above
upper limit is not much sensitive to the value of the exponent of the
power spectrum, $\gamma$. In this paper we are not going to discuss about
the derivation of the above constraint. Interested reader may
consult the reference we mentioned.
Now for the Coma cluster, the numerical 
value for the two parts of the scalar-to-photon conversion 
probability, eq.(\ref{sztemp}), turns out to be
\bea \label{sznu}
\frac {2  \langle\delta {\bf B}\cdot \delta {\bf B}\rangle }
{3{\cal N}^2 \omega^2} \varpi^2 \bar{\rho}_e^{-2} &\approx& \frac 
{2.7\times 10^{15}}{\omega^2}  \mbox{GeV}^2 \left(\frac {\delta
{\bf B}}{8.5 \mu\mbox{G}}\right)^2 \left(\frac {\varpi}{2 \pi ~204 \mbox{GHz}}
\right) \left(\frac {4\times10^{-3} \mbox{cm}^{-3}}{\rho_e}\right)^2, \nno\\ 
\frac {L_{eff} {\cal N}^{\gamma}{\cal P}_e^2 \langle\delta 
{\bf B}\cdot \delta {\bf B}\rangle }{6 |\gamma| \omega^2{\bar \rho}_{e}^2}
\varpi^{-\gamma} \bar{\rho}_e^{\gamma} &\approx&   \frac 
{2.44 \times 4.07^{\gamma} \times 10^{18+2\gamma}}{2 \pi \omega^2} 
\mbox{GeV}^2  \left(\frac {2 \pi ~204 \mbox{GHz}}
 {\varpi}\right)^{\gamma} \left(\frac {\rho_e}
{4\times10^{-3} \mbox{cm}^{-3}}\right)^{\gamma} \nno\\
&& ~~~~~~~~~~~~~~~~~~~~~~~~~~~~\times \left(\frac {\delta
{\bf B}}{8.5 \mu\mbox{G}}\right)^2  \left(\frac { L_{\bf B}}
{1 \mbox{kpc}}\right)^{\gamma+1}  \left(\frac { L}
{200 \mbox{kpc}}\right).  \nno \\
\eea  
From the above expressions, the constrain, on the 
scale of fine structure constant $\omega$, depends on several 
a priori unknown quantities like ICM magnetic field ${\bf B}$, 
electron density $\rho_e$, Coherence length $L_{\bf B}$, exponent
of the power spectrum of magnetic filed $\gamma$ etc. 
Accuracy of the constrains, therefore, depend severely on the 
observations of the properties of ICM. One can 
observe that depending upon the value of the power spectrum
exponent $\gamma$, the frequency dependence of ${\bar P}_{\gamma\rightarrow
\phi}$ changes which in turn affect on the upper bound of 
$\omega$. If we consider the aforementioned value of the power
spectrum exponent $-2<\gamma <-1$ and using the bound on 
$P^{coma}_{\gamma\rightarrow \phi}(204 \mbox{GeV})$ eq.\ref{szcoma}, 
one can get a contour plot Fig.(\ref{fig1}) for the 
parameter space $(\beta,y)$ where we denote $\omega = 10^y$. 
In the plot, we have considered the 
reasonable value of $I_e$ to be $\approx 1$ which is 
based on the electron density fluctuations in our own galaxy.
Numerical value of all the other parameters are considered to be
that of the aforementioned Coma galaxy cluster.
The shaded region is excluded due to non-observation of
any SZ-like effect coming from the scalar field. 

In order to elaborate
more on the possible bounds on the parameter space and also
compare with our previous results coming from the 
laboratory based experiments \cite{debupisin2}, in what
follows we consider a specific $\beta =2$ line in the parameter space
for two different values of $\gamma$.
It is straight forward to check that 
for $\beta \approx 2$, one gets
\bea \label{omegasP}
\omega^2 {\bar P}_{\gamma \rightarrow \phi} \approx 
(2.7-102.85)\times 10^{15}.
\eea
After using the above constrain coming from the Coma cluster Eq.\ref{szcoma}, 
for $\beta =2$ the above eq.\ref{omegasP} gives us the lower 
bound on $\omega$ to be
\bea \label{bound}
\omega \geq (0.66-4.04)\times 10^{10}~~ \mbox{GeV}.
\eea
It is clear from the above contour plot that as absolute value
of $\beta$ increases, the lower bound also increases for $\omega$.

\begin{figure}[t!]
\includegraphics[width=3.80in,height=3.50in]{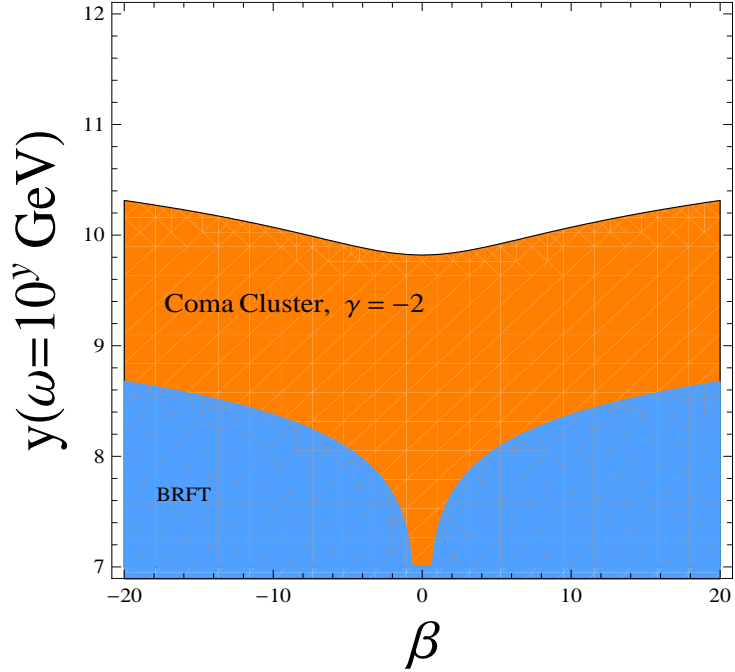}
\caption{\label{fig2} Comparing bounds on $\beta$ and $\omega$
coming from CMB observation on Coma galaxy cluster fig.\ref{fig1} 
and particular laboratory experiment
called BFRT from our previous paper \cite{debupisin2}}
\end{figure}

At this point it is worth comparing our present analysis with 
our previous bound coming from the laboratory based experiments we mentioned
before \cite{debupisin2}.
In our previous study we had considered different laboratory
based experimental results to constrain the model parameters.
Using the experimental constrain on the rotation and ellipticity of a
polarized electromagnetic wave passing through a 
magnetized region, we derived the
possible bounds to be 
$1 \leq \omega^2[\mbox{GeV}^2] \leq 10^{13} $ and
$-0.5 \leq \beta \leq 0.5$. The primary assumption behind those
constrains was that, $\beta$ should be less than unity. 
However, in our present study, it is clear that even if we consider
$\beta =0$, the parameter $\omega$ is always 
$ \gtrapprox \mbox{few}\times 10^{9} \mbox{GeV}$. 
It is also clear from the fig.\ref{fig2} that 
our previous bound on $\omega$ is completely excluded by the present 
CMB bound. Therefore, we can infer that if optical rotation and 
dichroism measured in the laboratory is sourced only by the possible 
variation of fine structure constant then it is almost impossible 
to see any positive signal with the current available values
of the experimental parameters and accuracy of the experiment.

Our present analysis does not help us
to constrain $\beta$. In order to constrain this 
we need to consider the polarization 
measurement of CMB. As we have mentioned before, with 
the present day experimental precession, it is very
difficult to measure the change of polarization due the
galaxy clusters. In the subsequent section we will discuss about
the prediction of our model on the change of the polarization of 
the CMB due to the ICM magnetized plasma. The future experiments
may shed some light on the existence of the 
parity violation through the CMB polarization measurement.

\section{Parity violating effect on CMB polarization} \label{sec4}

As we have already discussed before, in the presence of background 
ICM magnetic field and plasma density, varying alpha scalar
field alters the polarization states of the CMB photon. 
The leading order contribution to this change of the polarization 
comes from the photon-to-scalar conversion probability
$P_{\gamma \rightarrow \phi}$. The random part 
$P^{ran}_{\gamma \rightarrow \phi}$ contributes to the
stokes parameter ${\cal V}$ which gives rise to the 
circular polarization of the CMB photon. On the other hand, 
the regular part, $P^{reg}_{\gamma \rightarrow \phi}$, 
induces linear polarization of the CMB photon. The expressions
for the induced polarization of the CMB along the line of sight are
\bea
{\bar {\cal V}}(L) &\simeq& -\beta {\bar P}^{ran}_{\gamma\rightarrow \phi},\\
{\bar {\cal Q}}(L) &\simeq& {\bar P}^{reg}_{\gamma\rightarrow \phi}
 (\cos 2 \theta -
4 \beta \sin 2 \theta), \nno \\
{\bar {\cal U}}(L) & \simeq & {\bar P}^{reg}_{\gamma\rightarrow \phi}
 (\sin 2 \theta +
4 \beta \cos 2 \theta),
\eea
where the expression for the regular part of the conversion probability is 
\bea \label{preg}
{\bar P}^{reg}_{\gamma\rightarrow \phi}(L) =
\left( \frac {I_e^3 {\bf B}_{0}^2 }{{\cal N}^2 \omega^2} -
\frac {{\bf B}_0^2 \cos(\bar{\Delta} L_{eff})}{ {\cal N}^2 \omega^2}\right)
\varpi^2 \bar{\rho}_e^{-2} + \frac {\pi L_{eff} I_e^2 
{\cal N}^{\gamma}{\cal P}_e^2 {\bf B}_{0}^2}
{2 \gamma \omega^2{\bar \rho}_{e}^2}
\varpi^{-\gamma} \bar{\rho}_e^{\gamma}, \nno \\
\eea
and the expression for the random part is given in eq.\ref{sztemp}.
We have already discussed that the contribution from the
regular magnetic field part ${\bf B}_0$ is very small compared 
to the contribution from the random magnetic field $\delta {\bf B}$.
To the leading order in $1/\omega^2$, therefore, magnitude of the
induced circular polarization ${\bar V}(L) \gg {\bar Q}(L), 
{\bar  U}(L)$. It is easy to see that 
the dominant contribution 
in the Eq.\ref{preg} is coming from the electron density fluctuation. 
Using all the measured quantity for the Coma cluster, one gets
\bea
 \frac {\pi L_{eff} {\cal N}^{\gamma}{\cal P}_e^2 
{\bf B}_0^2 }{2 |\gamma| \omega^2{\bar \rho}_{e}^2}
\varpi^{-\gamma} \bar{\rho}_e^{\gamma} &\approx&   \frac
{6.4 \times 4.07^{\gamma} \times 10^{14+2\gamma}}{\omega^2}
\mbox{GeV}^2  \left(\frac {2 \pi ~204 \mbox{GHz}}
 {\varpi}\right)^{\gamma} \left(\frac {\rho_e}
{4\times10^{-3} \mbox{cm}^{-3}}\right)^{\gamma} \nno\\
&& ~~~~~~~~~~~~~~~~~~~~~~~~~~~~\times \left(\frac {\delta
{\bf B}}{0.2 \mu\mbox{G}}\right)^2  \left(\frac { L_{\bf B}}
{1 \mbox{kpc}}\right)^{\gamma+1}  \left(\frac { L}
{200 \mbox{kpc}}\right).   
\eea
This is clearly in magnitude of the order of $10^{-4}$ lower than that of 
${\bar P}^{ran}_{\gamma\rightarrow \phi}(L)$ mentioned before.
If we use our previous bound on $\omega$ from the Eq.\ref{bound},
the prediction of the linear polarization, which is coming from the 
regular part of the magnetic field in the ICM, comes out to be 
\bea
{\bar P}^{reg}_{\gamma\rightarrow \phi}(L) \leq (0.89-1.51)\times 10^{-10}.
\eea
This is much less than that of the circular polarization which
is proportional to 
\bea
{\bar P}^{ran}_{\gamma\rightarrow \phi}(L) \leq 6.2 \times 10^{-5},
\eea
The interesting point, we would like mention, is about its 
non-trivial dependence
on the parity violating parameter $\beta$. We have mentioned before
that $\beta$ could be greater than unity. Because of this fact, two
linear polarizations ${\bar {\cal Q}}(L)$ and ${\bar {\cal Q}}(L)$
could be of opposite sign. In principle this effect can 
potentially be detectable from the next generation experiments.  
There exist few earth based experiments such as SPT-Pol, ALMA, POLAR, 
which are either ongoing or under development, 
have detectors to measure the polarization of CMB also on the low scale.
All these experiments with high angular resolution could in principle 
shed some light on the parity violating effect on the 
polarization of the CMB.  

Before closing this section we would like to discuss about the 
induced polarization coming from the random magnetic component of 
the ICM. We have seen that the circular polarization is induced by the 
random part of the magnetic field and the linear
polarization can also get some contribution from the random
part of the magnetic field. Since the magnitude and direction of the 
polarization stokes parameters coming from the random
part of the ICM magnetic field depend on its magnitude and direction
, the average over the many line of sights of those random 
contributions will vanish. The effective contribution from the 
random magnetic field, therefore, can be encoded in the variance 
$\sigma^2$ of those stokes parameters ${\cal Q}, {\cal U} $ and ${\cal V}$.
Now according to the standard definition, one can get the variance
of the stokes parameters ${\cal Q}$ to be \cite{shaw},    
\bea
\sigma_{\cal Q}^2 &=& \langle {\cal Q}(\Delta_1) {\cal Q}(\Delta_2)\rangle
- \langle {\cal Q}(\Delta_1)\rangle \langle{\cal Q}(\Delta_2)\rangle, \nno \\
&\approx& \frac 1 2 \left( 
P^2_{\gamma \rightarrow \phi}(\Delta_1,\Delta_2) - \beta^2 
{P^{ran}_{\gamma \rightarrow \phi}}^2 (\Delta_1,\Delta_2)
 + {P^{reg}_{\gamma \rightarrow \phi}}^2(\Delta_1,\Delta_2)
((1-16 \beta^2)\cos 4 \theta - 8 \beta \sin 4 \theta )\right) \nno \\
&&~~~~~~~~~~~~~~~~~~~~~~~~~~~+ (\Delta_2\rightarrow -\Delta_2).\nno
\eea
In the above derivation we have approximated ${\cal B}'s$ to be
Gaussian and therefore all the expectation values can be written
in terms of two point correlation function.  
With the similar definition, it can be easily shown that the variance of 
the other stokes parameters come out to be:
\bea
\sigma_{\cal U}^2 &\approx& \frac 1 2 \left(   
P^2_{\gamma \rightarrow \phi}(\Delta_1,\Delta_2) - \beta^2
{P^{ran}_{\gamma \rightarrow \phi}}^2 (\Delta_1,\Delta_2) 
- {P^{reg}_{\gamma \rightarrow \phi}}^2(\Delta_1,\Delta_2)
((1-16 \beta^2)\cos 4 \theta - 8 \beta \sin 4 \theta)\right) \nno \\
&&~~~~~~~~~~~~~~~~~~~~~~~~~~~~~~+ (\Delta_2\rightarrow -\Delta_2), \nno \\
\sigma_{\cal V}^2 &\approx& \frac 1 2 \left(   
P^2_{\gamma \rightarrow \phi}(\Delta_1,\Delta_2) + \beta^2
{P^{ran}_{\gamma \rightarrow \phi}}^2 (\Delta_1,\Delta_2) 
- {P^{reg}_{\gamma \rightarrow \phi}}^2(\Delta_1,\Delta_2)
(1 + 4\beta^2)\right) + 
(\Delta_2\rightarrow -\Delta_2). \nno
\eea
As we have discussed in the previous section, the contribution
from the random magnetic field is much greater than the 
regular contribution in the photon to scalar conversion probability.
So to the leading order in magnitude we can clearly see 
from the above variance that all the polarization Stokes parameters
are proportional to the $P_{\gamma \rightarrow \phi}^{ran}$.
The expression for the variance can be approximated as
\bea
\sigma_{\cal Q}^2 &\approx& \frac 1 2 
{P^{2}_{\gamma \rightarrow \phi}} (\Delta_1,\Delta_2)  - \frac {\beta^2}2
{P^{ran}_{\gamma \rightarrow \phi}}^2 (\Delta_1,\Delta_2 )+ 
(\Delta_2\rightarrow -\Delta_2),\nno \\
\sigma_{\cal U}^2 &\approx& \frac 1 2
{P^{2}_{\gamma \rightarrow \phi}} (\Delta_1,\Delta_2)  - \frac {\beta^2} 2
{P^{ran}_{\gamma \rightarrow \phi}}^2 (\Delta_1,\Delta_2 )+
(\Delta_2\rightarrow -\Delta_2),\nno \\
\sigma_{\cal V}^2 &\approx& \frac 1 2
{P^{2}_{\gamma \rightarrow \phi}} (\Delta_1,\Delta_2)  +\frac {\beta^2} 2
{P^{ran}_{\gamma \rightarrow \phi}}^2 (\Delta_1,\Delta_2 )+
(\Delta_2\rightarrow -\Delta_2).\nno 
\eea
But as we have stated before, experimentally the random contribution is
very difficult to measure with the present level of experimental accuracy. 
Regarding this problems of measurement, an
elaborate discussion has been provided in reference \cite{shaw}.
We are not going to discuss it further. 
The essential point, that we would like to infer, is that for the 
contribution coming from the regular magnetic field part ${\bf B}_0$
of the ICM, we do not have such measurement problems. Although
the magnitude of that contribution ($\sim 10^{-10}$) is 
very small compared to the intrinsic polarization of the CMB photon 
($\sim 10^{-7}$), the recent experiment like ALMA, with
the order of few arc second angular resolution, 
could help to put stringent bound on our model parameters or 
in principle could detect some positive signal regarding the parity 
violation in the photon sector.  

\section{Conclusions} \label{con} 
The theory of varying fine structure constant has been the
subject of intense study
in the last several years. Cosmological impact of this
variation has been studied quite extensively. Various cosmological
as well as laboratory based observations on this variation
of fine structure constant have been considered to constrain
the varying alpha parameter
$\omega$. Recently we have constructed a particular model based on
this varying alpha theory which includes explicit PCP
violation in the photon sector \cite{debupisin}.
In this paper we have studied our aforementioned PCP violating
varying alpha model in the light of a new class of
cosmological observations which have
not been considered before.
We considered the SZ measurement of Coma galaxy cluster 
to constrain our model parameters. 
In this particular measurement the temperature variation of CMB
is being measured. The basic underlying mechanism behind 
this measurements is the existence of a non-trivial 
interaction between photon and
high temperature plasma field in the ICM. As stated before, if there
exists a light scalar field which has non-trivial
coupling with the photon field then one would expect additional
SZ-like effect on CMB. This is what we have studied in detail
in this paper. In our model \cite{debupisin}
we have introduced a non-trivial PCP violating scalar-photon
interaction within the varying alpha theory framework. Although
the experiments under consideration are
insensitive to the properties of the background field
due to the weakness of its coupling with matter,
they nevertheless can help to constrain our varying alpha model parameters
$\omega$ and $\beta$ through the possible frequency-dependent 
upper limit on the temperature variation within the error bar 
of the usual thermal SZ measurement. 
We have calculated the approximate analytic expression of our model 
for those measurable quantities such as stokes parameters of CMB photon
passing through the ICM.
The model is characterized by two independent parameters
$\beta$ and $\omega$ that measure the strength of
PCP violation and the scale of fine structure constant variation,
respectively. 

As we mentioned before in our previous study \cite{debupisin2} 
we had considered different laboratory
based experimental results to constrain our model parameters.
In our present study we use the SZ measurement of CMB photon passing 
through the galaxy cluster to constrain our parameters. Using the 
measurement on Coma galaxy cluster we found from fig.\ref{fig1},
the lower bound on $\omega$ depends on the value of PCP 
violating parameter $\beta$. If we choose $\beta=0$ line which
corresponds the standard Maxwell-dilaton type model, we 
approximately reproduce the known bound 
$\omega \geq 10^{9} \mbox{GeV}$. According to our study in 
this paper the polarization measurement
of CMB photon is essential to constrain the parity violating parameter.
If the fine structure constant is varying then the variation can
lead to a certain degree of linear and circular polarizations to
the CMB photon when it is passing through the magnetized medium.
We have a definite prediction on the amount of circular polarization
and linear polarization. But short-coming is that even though 
the circular polarization is induced by the parity 
violating parameter $\beta$, 
the contribution is coming from the random magnetic field part of ICM.
As has been mentioned, it is very difficult to detect 
this signal mainly because of its random nature over a very small
angular scale. In other words, the line of sights
are typically separated by distance of the order of the coherence scale
($L_{\bf B}$) of random magnetic field. In order to detect the signal,
one, therefore, needs to increase the angular resolution of an experiment
to a very high precession.  
There exist measurements of polarization of photons at the galactic scale
such as that of the Milky Way. We could in principle use those
measurements to constrain our model parameter and also check 
the consistency with present bound. 
 
\vspace{.1cm}

{\bf Acknowledgement}\\
Referee's valuable comments and suggestions are gratefully 
acknowledged. This research is supported by 
Taiwan National Science Council under Project No. NSC
97-2112-M-002-026-MY3, by Taiwan's National Center
for Theoretical Sciences (NCTS), and by US Department
of Energy under Contract No. DE-AC03-76SF00515.

\section*{Appendix}
In this section we will provide all the details of the calculation
for the stokes parameters. In order to get the expression for 
the photon-to-scalar conversion probability amplitude and the polarization 
of state we consider the following procedure. 
We divide the total magnetic field as a regular 
and a random part like ${\bf B}= {\bar {\bf B}} + \delta {\bf B}$. 
Similarly we can define the
total electron density as $\rho_e = {\bar \rho}_e + \delta \rho_e$  
where ${\bar \rho}_e$ is the constant average electron density 
over the galaxy cluster $L$.
Let us define a new quantity 
\bea
\overline {{{\cal D} {\bf B}^2}}(z) &=& 
\frac  1 z \int_0^z (2 {\bar {\bf B}}\cdot \delta {\bf B}(x) + 
\delta {\bf B}(x) \cdot \delta {\bf B}(x)) dx =
\frac  1 z \int_0^z {{\cal D} {\bf B}^2}(x) dx  \\
{\bar {\delta_e}}(z)& =&  \frac  1 z \int_0^z {\delta_e(x)} dx= \frac  1 z \int_0^z \frac {\delta \rho_e(x)}
{{\bar \rho}_e} dx,
\eea
such that due to randomness of the density fluctuation over the length L, 
${\bar {\delta_e}}(L) =0$.
If we define a new variable $Z= (1 + \delta_e)x $,  the integral 
\bea
{\cal B}_i = \int_0^L ({\bf B}_i -2 \beta \epsilon^{ij} {\bf B}_j) e^{i S} dx
= \int_0^L \frac {({\bf B}_i -2 \beta \epsilon^{ij} {\bf B}_j)}
{1 + \delta_e} 
e^{- i \left({\bar \Delta} + \frac 
{\overline {{{\cal D} {\bf B}^2}}(Z)} {\varpi \omega^2} 
\right)Z}  dZ,
\eea
where
\bea
{\bar \Delta}  = \left(\frac {2 \pi \alpha_{0} {\bar \rho}_e}
{2 m_e \varpi} - \frac
{{\bar {\bf B}}^2} {\varpi \omega^2}\right).  
\eea
In the above expressions we assume $|{\bar {\delta_e}}(z)|\ll 1$ along the 
photon path. For further simplification it would be useful to 
do another change of variable like 
\bea
T = \left(1 + \frac 
{\overline {{{\cal D} {\bf B}^2}}(Z)} {\varpi \omega^2 {\bar \Delta}} 
\right)Z.
\eea
Therefore, the final expression for ${\cal B}_i$ takes the form 
\bea
{\cal B}_i({\bar \Delta}) = \int_0^{L_{eff}} \frac 
{({\bf B}_i -2 \beta \epsilon^{ij} {\bf B}_j)}
 { \left(1 + \frac         
{ {{{\cal D} {\bf B}^2}}(T)} {\varpi \omega^2 {\bar \Delta}} 
\right) (1 + \delta_e)}
e^{- i {\bar \Delta} T}  dT,
\eea 
where $L_{eff} = \left(1 + \frac         
{\overline {{{\cal D} {\bf B}^2}}(L)} {\varpi \omega^2 {\bar \Delta}} 
\right)L$.
As we have mentioned before to estimate the 
intensity $I(z)$ and polarization states 
${\cal Q}(z), {\cal U}(z)$ and ${\cal V}(z)$ of the CMB photon, the main 
quantity of our interest is $G_{ij}$. In term of the 
new variable as explained above, $G_{ij}$ takes 
the following form at  different frequencies
\bea
G_{ij}({\bar \Delta},{\bar \Delta}') =
< {\cal B}^*_i(T,{\bar \Delta}) {\cal B}_j(T',{\bar \Delta}')> 
= \int_0^{L_{eff}}\int_0^{L_{eff}} {\cal F}_{ij}(T,T')
e^{ i ({\bar \Delta}  T - {\bar \Delta}'  T' )}  dT dT'.
\eea  
We define the correlation function as 
\bea \label{corr}
{\cal F}_{ij}(T,T') =
\Bigg\langle \frac {({\bf B}_i -2 \beta \epsilon^{ik} {\bf B}_k)
({\bf B}'_i -2 \beta \epsilon^{il} {\bf B}'_l)}
 { \left(1 + \frac
{ {{{\cal D} {\bf B}^2}}(T)} {\varpi \omega^2 {\bar \Delta}}
\right) \left(1 + \frac
{ {{{\cal D} {\bf B}^2}}(T')} {\varpi' \omega^2 {\bar \Delta}'}
\right)} \Bigg \rangle \Big\langle \frac 1 {(1 + \delta_e)(1 + \delta_e')} 
\Big\rangle
\eea
In the above expression for the correlation function we assume that 
the magnetic field fluctuation $\delta {\bf B}$ and electron density
fluctuation $\delta \rho_e$ are uncorrelated. Isotropy of the fluctuations
can simplify the above Eq.\ref{corr} for ${\cal F}_{ij}(T,T')$ to
\bea
{\cal F}_{ij}(T,T') =
\Bigg \langle \frac {{\bar {\bf B}}^{eff}_i {\bar {\bf B}}^{eff}_j}
 { \left(1 + \frac
{ {{{\cal D} {\bf B}^2}}(T)} {\varpi \omega^2 {\bar \Delta}}
\right) \left(1 + \frac
{ {{{\cal D} {\bf B}^2}}(T')} {\varpi' \omega^2 {\bar \Delta}'}
\right)}  +
 \frac 1 3  \frac { {\cal E}_{ij}\delta{\bf B} \cdot 
\delta{\bf B}'}
 { \left(1 + \frac
{ {{{\cal D} {\bf B}^2}}(T)} {\varpi \omega^2 {\bar \Delta}}
\right) \left(1 + \frac
{ {{{\cal D} {\bf B}^2}}(T')} {\varpi' \omega^2 {\bar \Delta}'}
\right)} \Bigg \rangle  R_{\delta}(T,T').
\eea
Where we have defined 
\bea
&&{\cal E}_{ij} = (\delta_{ij} - 4 \beta \epsilon_{ij} + 4 \beta^2 \epsilon_{ik}
\epsilon_{jk}) ~~;~~{\bar {\bf B}}^{eff}_i = 
({\bar {\bf B}}_i -2 \beta \epsilon^{ik} {\bar {\bf B}}_k) \nno\\
&&R_{\delta}(T,T') = \Big\langle \frac 1 {(1 + \delta_e)(1 + \delta_e')}
\Big\rangle
\eea
We have assumed the isotropy and the approximate position independent
fluctuations of the cluster magnetic field and electron density. 
With this assumption the power spectrum of physical interests 
can then be defined by a simple Fourier transformation
\bea
G_{ij}({\bar \Delta},{\bar \Delta}') =
= \frac 1 {4 \pi} \int_0^{L_{eff}} dT \int_0^{L_{eff}} dT' \int d^3 k
{\cal F}^k_{ij} e^{i k(T-T')}
e^{ i ({\bar \Delta}  T - {\bar \Delta}'  T' )} ,
\eea
In order to proceed further we will do some approximation adopting from 
\cite{shaw}. We have mentioned before that we will be interested
in dealing with the CMB photon passing through the ICM. 
The typical frequency of the CMB photon is $\varpi \simeq
10^{-5}-10^{-3} eV$ propagating over the distance around 100 kpc through the
galaxy clusters. One can easily check therefore that in general 
$|{\bar \Delta}| L_{eff} \gg 1$ as long as 
$ {2 \pi \alpha_{0} {\bar \rho}_e}/
{ m_e }$ is not finely tuned to be $\simeq 
{{\bar {\bf B}}^2}/{\omega^2} $. In order to get an approximate analytic
expression for the above correlation function $G_{ij}(T,T')$,
we further assume that the fluctuation of ${\bf B}$ and $\rho_e$  
are such that ${\cal F}^k_{ij}$ falls off faster that $k^{-3}$ for
$k>k_*$. Where $k_*^{-1}$ should be related to the characteristic 
coherent lengths $L_{\bf B}$ and $L_e$ of 
${\bf B}$ and $\rho_e$ fluctuations respectively in a 
galaxy cluster under consideration. 
This is also believed to be a reasonable assumption that 
max$({\bar \Delta},{\bar \Delta}')\ll L_{\bf B}^{-1}, L_e^{-1}$.
With all these assumptions and considering 
max$(|{\bar \Delta}| L_{eff},|{\bar \Delta}'| L_{eff}) \gg 1$
one can get the following expression to the leading order
\bea \label{gij}
e^{-i(\Delta_- L_{eff})} G_{ij}({\bar \Delta},{\bar \Delta}')\approx 
\frac {2 \cos(\Delta_- L_{eff})}{{\bar \Delta}{\bar \Delta}'} 
{\cal F}_{ij}(0) &-& \frac {2 \cos(\Delta_+ L_{eff})}
{{\bar \Delta}{\bar \Delta}'} 
{\cal F}_{ij}(L_{eff} 
\hat{{\bf z}})  \\
&+& \frac {\pi \sin(\Delta_- L_{eff})}{2\Delta_- }\left[ 
\int_{{\bar \Delta}}^{\infty} k dk {\cal F}^k_{ij} + 
\int_{{\bar \Delta}'}^{\infty} k dk
{\cal F}^k_{ij} 
\right] \nno
\eea
where $\Delta_{\pm} = ({\bar \Delta} \pm {\bar \Delta}')/2$. $\hat{{\bf z}}$
is the direction along the propagation of light.
In the above expression for the two point correlation function
\bea
{\cal F}_{ij}(0) =
\Bigg \langle \frac {{\bar {\bf B}}^{eff}_i {\bar {\bf B}}^{eff}_j}
 { \left(1 + \frac
{ {{{\cal D} {\bf B}^2}}} {\varpi \omega^2 {\bar \Delta}}
\right)^2 }  +
 \frac 1 3  \frac { {\cal E}_{ij}\delta{\bf B} \cdot
\delta{\bf B}}
 { \left(1 + \frac
{ {{{\cal D} {\bf B}^2}}} {\varpi \omega^2 {\bar \Delta}}
\right)^2 } \Bigg \rangle  \Big \langle \frac {{\bar \rho}^2}{\rho^2} \Big 
\rangle ,
\eea
where we have used the relation 
$\varpi' \Delta' = \varpi \Delta$.
The effective distance $L_{eff}$, through which the CMB photon is traveling,
is much larger than the coherence length of the fluctuations. So leading 
contribution to ${\cal F}_{ij}(L_{eff})$  should be coming from the regular
magnetic field part of the ICM.
\bea
{\cal F}_{ij}(L_{eff})=
\Bigg \langle \frac {{\bar {\bf B}}^{eff}_i {\bar {\bf B}}^{eff}_j}
 { \left(1 + \frac
{ {{{\cal D} {\bf B}^2}}(x)} {\varpi \omega^2 {\bar \Delta}}
\right) \left(1 + \frac
{ {{{\cal D} {\bf B}^2}}(x+L_{eff})} {\varpi \omega^2 {\bar \Delta}}
\right)}  \Bigg \rangle  
\eea
Now we need to express last two terms of Eq.\ref{gij} in terms of known
correlation functions. If we assume the fluctuation of $\rho_e$ is
log-normal then one can write down
\bea
\Big \langle \frac {{\bar \rho}^2}{\rho^2} \Big
\rangle = \Big \langle \frac {\rho^2}{{\bar \rho}^2}\Big
\rangle^3 = I_e^3
\eea 
Based on this log-normal distribution, it is consistent to separate  
the fluctuation $\rho_e$ into approximately independent
short and long wavelength fluctuation such that $\rho_e = {{\bar \rho}_e} 
(1+ \delta_s)(1+\delta_l)$. We also assume that the short wavelength 
fluctuations are linear up to some cut-off scale $k^{-1}_{lin}$.
The long wave length fluctuations are assumed to be above this scale
and not necessarily be linear. With the above assumptions, one can
show \cite{shaw}
\bea
R_{\delta} \simeq I_{e}^2(x)[1+ {{\bar \rho}_e}^{-2} R_e(x)].
\eea
From the above expression for the momentum $k\gg k_{lin}$, we can
approximately have
\bea
R^k_{\delta} =  I_{e}^2 {{\bar \rho}_e}^{-2} P_e(k)].
\eea

Now, let us define
\bea
\int_{{\bar \Delta}}^{\infty} k dk {\cal F}^k_{ij} ={\bar {\bf B}}^{eff}_i 
{\bar {\bf B}}^{eff}_j \int_{{\bar \Delta}}^{\infty} k dk {\cal F}^k_{(0)}
+ \frac {{\cal E}_{ij}} 3 \int_{{\bar \Delta}}^{\infty} 
k dk {\cal F}^{k}_{(1)}. 
\eea  
Following argument in \cite{shaw}, we can write down
\bea
\int_{{\bar \Delta}}^{\infty} k dk {\cal F}^k_{(0)}
&=& \Bigg \langle \frac {1}  { \left(1 + \frac
{ {{{\cal D} {\bf B}^2}}} {\varpi \omega^2 {\bar \Delta}} \right)^2 }  
\Bigg \rangle
\int_{{\bar \Delta}}^{\infty} k dk {R_{\delta}}^k
+ I_e^3 \int_{{\bar \Delta}}^{\infty} k dk {R^k_{0{\bf B}}}, \\
\int_{{\bar \Delta}}^{\infty} k dk {\cal F}^k_{(1)}
&=& \Bigg \langle \frac {\delta {\bf B}\cdot \delta {\bf B}}  { \left(1 + \frac
{ {{{\cal D} {\bf B}^2}}} {\varpi \omega^2 {\bar \Delta}} \right)^2 }   
\Bigg \rangle
\int_{{\bar \Delta}}^{\infty} k dk {R_{\delta}}^k
+ I_e^3 \int_{{\bar \Delta}}^{\infty} k dk {R^k_{1{\bf B}}},
\eea
where new definitions are
\bea
\Bigg \langle \frac {1}
 { \left(1 + \frac
{ {{{\cal D} {\bf B}^2}}(T)} {\varpi \omega^2 {\bar \Delta}}
\right) \left(1 + \frac
{ {{{\cal D} {\bf B}^2}}(T')} {\varpi' \omega^2 {\bar \Delta}'}
\right)} \Bigg \rangle = \frac 1 {4 \pi} \int  d^3 k R^k_{0{\bf B}} 
e^{i(T-T')\cdot k} \\
\Bigg \langle\frac { \delta{\bf B} \cdot
\delta{\bf B}'}
 { \left(1 + \frac
{ {{{\cal D} {\bf B}^2}}(T)} {\varpi \omega^2 {\bar \Delta}}
\right) \left(1 + \frac
{ {{{\cal D} {\bf B}^2}}(T')} {\varpi' \omega^2 {\bar \Delta}'}
\right)} \Bigg \rangle = \frac 1 {4 \pi} 
\int d^3 k R^k_{1{\bf B}} e^{i (T-T')\cdot k}
\eea
Therefore, all the physical quantities, 
Eq.\ref{pquv} that we are interested in, can 
be expressed  in terms of $G_{ij}$ as follows:
\bea
P_{\gamma\rightarrow \phi} &=&  \frac 1 {2 \omega^2} \delta^{ij} G_{ij} ~~~;~~~
{\cal Q}(z) = -\frac {1}{2 \omega^2} (\delta^{xi}\epsilon^{yj} +
\delta^{yi}\epsilon^{xj}) G_{ij}, \nno \\
{\cal U}(z) & = & \frac {1}{2 \omega^2} (\delta^{xi}\epsilon^{xj} -
\delta^{yi}\epsilon^{yj}) G_{ij}  ~~~;~~~{\cal V}(z) =  
\frac {1}{2 \omega^2} \epsilon^{ij}  G_{ij},
\eea
where index "i" is running for $x$ and $y$ coordinate. 
It is straight forward to check from the above expressions that,
$V(z)$ is non-vanishing as expected from the PCP violating term. 
Therefore, to the leading order, it should be proportional to 
$\beta/\omega^2$. By using the following identities, 
$\delta^{ij} {\cal E}_{ij} = \delta^{i}_{i}(1 + 4 \beta^2)$
and $\epsilon^{ij} {\cal E}_{ij} = - 4 \beta \delta^{i}_{i}$, one
gets
\bea
{\bar V}(L) \simeq -\frac {2 \beta I(0) e^{-i(\Delta_- L_{eff})}}
{3\omega^2}\left[   
\frac {2 \cos(\Delta_- L_{eff}) I_e^3 }
{ {\bar \Delta}{\bar \Delta}' } 
\right. && \left. \Bigg \langle  \frac { \delta{\bf B} \cdot
\delta{\bf B}} { \left(1 + \frac
{{\cal D} {\bf B}^2} {\varpi \omega^2 {\bar \Delta}}
\right)^2 } \Bigg \rangle  \right. \\
&& \left. + 
\frac { \pi \sin(\Delta_- L_{eff})}{\Delta_- }\left(
\int_{{\bar \Delta}}^{\infty} + \int_{{\bar \Delta}'}^{\infty}\right)
k dk {\cal F}^k_{(1)} \right]. \nno
\eea

The PCP violating term in our Lagrangian, therefore, induces 
a certain degree of circular polarization to the incoming 
CMB photon propagating through the ICM magnetized plasma. 
The contribution is strongly depending
upon the fluctuating part of the ICM magnetic field.   
If we have only regular part of the magnetic field, the induced 
circular polarization vanishes to the leading order in
PCP violating parameter $\beta$.

If we consider the regular component of the ICM magnetic 
field $\bar{\bf B}_x = {\bf B}_0 \cos \theta$ and 
$\bar{\bf B}_y = {\bf B}_0 \sin \theta$, then one can easily show 
the following approximate expressions for the other stokes parameters
\bea
{\bar {\cal Q}}(L) &=& -\frac {1}{2 \omega^2} (\delta^{xi}\epsilon^{yj} +
\delta^{yi}\epsilon^{xj}) G_{ij} \simeq e^{-i(\Delta_- L_{eff})}
{\cal A}({\bar \Delta},{\bar \Delta}')
 (\cos 2 \theta - 
4 \beta \sin 2 \theta), \nno \\
{\bar {\cal U}}(L) & = & \frac {1}{2 \omega^2} (\delta^{xi}\epsilon^{xj} -
\delta^{yi}\epsilon^{yj}) G_{ij}  \simeq e^{-i(\Delta_- L_{eff})}
{\cal A}({\bar \Delta},{\bar \Delta}')
 (\sin 2 \theta +   
4 \beta \cos 2 \theta),
\eea
where the expression for ${\cal A}$ is
\bea \label{Aex}
\omega^2 {\cal A} ({\bar \Delta},{\bar \Delta}')\approx 
&+& \frac { I_e^3 {\bf B}_0^2 \cos(\Delta_- L_{eff})}
{{\bar \Delta}{\bar \Delta}'} \Bigg \langle \frac {1}  { \left(1 + \frac
{ {{{\cal D} {\bf B}^2}}} {\varpi \omega^2 {\bar \Delta}} \right)^2 }  
\Bigg \rangle \nno \\ 
&-& \frac { {\bf B}_0^2 \cos(\Delta_+ L_{eff})}{{\bar \Delta}{\bar \Delta}'} 
\Bigg \langle \frac {1}
 { \left(1 + \frac
{ {{{\cal D} {\bf B}^2}}(x)} {\varpi \omega^2 {\bar \Delta}}
\right) \left(1 + \frac
{ {{{\cal D} {\bf B}^2}}(x+L_{eff})} {\varpi \omega^2 {\bar \Delta}}
\right)}  \Bigg \rangle \nno \\
&+& \frac {\pi {\bf B}_0^2 \sin(\Delta_- L_{eff})}{4\Delta_- }\left[ 
\int_{{\bar \Delta}}^{\infty} k dk {\cal F}^k_{(0)} + 
\int_{{\bar \Delta}'}^{\infty} k dk
{\cal F}^k_{(0)} 
\right]. \nno
\eea
Due to the PCP violation, two linear polarization states of the CMB are
effected oppositely to the leading order in $\beta/\omega^2$. 
Finally the expression for the photon-to-scalar conversion probability
amplitude becomes
\bea
e^{i(\Delta_- L_{eff})}{\bar P}_{\gamma\rightarrow \phi} 
&=&  \frac 1 {2 \omega^2}e^{i(\Delta_- L_{eff})} \delta^{ij} G_{ij}\nno\\
& \approx& 
+ \frac { I_e^3 \cos(\Delta_- L_{eff})}
{{\bar \Delta}{\bar \Delta}' \omega^2} 
\Bigg \langle \frac {{\bf B}_0^2+ \frac 2 3
\delta{\bf B}\cdot \delta {\bf B} }  { \left(1 + \frac
{ {{{\cal D} {\bf B}^2}}} {\varpi \omega^2 {\bar \Delta}} \right)^2 }  
\Bigg \rangle \nno \\ 
&& - \frac { {\bf B}_0^2 \cos(\Delta_+ L_{eff})}{{\bar \Delta}{\bar \Delta}'
\omega^2} 
\Bigg \langle \frac {1}
 { \left(1 + \frac
{ {{{\cal D} {\bf B}^2}}(x)} {\varpi \omega^2 {\bar \Delta}}
\right) \left(1 + \frac
{ {{{\cal D} {\bf B}^2}}(x+L_{eff})} {\varpi \omega^2 {\bar \Delta}}
\right)}  \Bigg \rangle  \\
&& + \frac {\pi \sin(\Delta_- L_{eff})}{4\Delta_- \omega^2}\left[ 
{\bf B}_0^2 \left(\int_{{\bar \Delta}}^{\infty} + 
\int_{{\bar \Delta}'}^{\infty}\right)  k dk {\cal F}^k_{(0)} + 
\frac {2(1+4\beta^2)} 3 \left(\int_{{\bar \Delta}}^{\infty} + 
 \int_{{\bar \Delta}'}^{\infty}\right) k dk {\cal F}^k_{(1)} 
\right]. \nno 
\eea
It is important to note that, the photon-to-scalar 
conversion probability depends on the parity 
violating parameter to the order $\beta^2$. 
The intuitive reason behind this is that the energy density
of the electromagnetic field does not depend on $\beta$ linearly. 
In order to express all the above quantities in terms of magnetic and
electron density power spectrum, we need to use perturbative  expansion.
To the leading order in $1/{\omega^2}$, all the correlation functions
appeared in the above expressions for the stokes parameters can
be expressed as follows:
\bea
&&\Bigg \langle \frac {1}  { \left(1 + \frac
{ {{{\cal D} {\bf B}^2}}} {\varpi \omega^2 {\bar \Delta}} \right)^2 }
\Bigg \rangle 
\approx \Bigg \langle 1  -2 \frac
{ {{{\cal D} {\bf B}^2}}} {\varpi \omega^2 {\bar \Delta}}  
\Bigg \rangle = 1  -2 \frac
{ \langle \delta  {\bf B} \cdot \delta {\bf B}\rangle} {\varpi \omega^2 
{\bar \Delta}},  \nno \\
&&\Bigg \langle \frac {1}
 { \left(1 + \frac
{ {{{\cal D} {\bf B}^2}}(x)} {\varpi \omega^2 {\bar \Delta}}
\right) \left(1 + \frac
{ {{{\cal D} {\bf B}^2}}(x')} {\varpi \omega^2 {\bar \Delta}}
\right)}  \Bigg \rangle \approx  1  -2 \frac
{\langle \delta  {\bf B} \cdot \delta {\bf B}\rangle} {\varpi \omega^2 
{\bar \Delta}},  \nno \\
&&\Bigg \langle \frac {\delta {\bf B}(x) \cdot \delta {\bf B}(x')}
 { \left(1 + \frac
{ {{{\cal D} {\bf B}^2}}(x)} {\varpi \omega^2 {\bar \Delta}}
\right) \left(1 + \frac
{ {{{\cal D} {\bf B}^2}}(x')} {\varpi \omega^2 {\bar \Delta}}
\right)}  \Bigg \rangle \approx  \langle 
\delta {\bf B}(x) \cdot \delta {\bf B}(x')\rangle
\left(1 - \frac {10} 3  \frac{ \langle \delta  {\bf B} \cdot \delta 
{\bf B}\rangle} {\varpi \omega^2 
{\bar \Delta}} \right) .
\eea
In the above expressions, we assume that the distribution of 
the fluctuating magnetic field $\delta
{\bf B}$ in the ICM is approximately gaussian. 
With this approximation, one can express all
the unknown correlation functions in terms of ICM magnetic and 
electron density power spectrum namely $P_{\bf B}$ and $P_e$ as 
follows:
\bea
\int_{{\bar \Delta}}^{\infty} k dk {\cal F}^k_{(0)}
&=&   I_{e}^2 {{\bar \rho}_e}^{-2}\left(1 -2 \frac
{ \langle  \delta {\bf B}\cdot \delta{\bf B}\rangle} 
{\varpi \omega^2 {\bar \Delta}} \right)   
 \int_{{\bar \Delta}}^{\infty} k dk {P_{e}}(k) , \\
\int_{{\bar \Delta}}^{\infty} k dk {\cal F}^k_{(1)}
&=&\left(1 - \frac {10} 3 \frac{ \langle \delta  {\bf B} \cdot \delta 
{\bf B}\rangle} {\varpi \omega^2 {\bar \Delta}} \right)
\left[ I_{e}^2 {{\bar \rho}_e}^{-2}\langle \delta  {\bf B} \cdot \delta
{\bf B}\rangle \int_{{\bar \Delta}}^{\infty} 
k dk {P_{e}}(k) + I_e^3 \int_{{\bar \Delta}}^{\infty} 
k dk {P_{\bf B}}(k)\right].\nno 
\eea
It is important to note that all the relevant
observable quantities are depending on
a critical length scale called ${\bar \Delta}^{-1}$.

\end{document}